\documentclass[showpacs ,showkeys,twocolumn]{revtex4}
\usepackage{amsmath,amsfonts,paralist}
\usepackage{graphicx}
\usepackage{epstopdf}
\usepackage{bm}

\begin{document}

\title{A stroll among effective temperatures 
in aging systems:
limits and perspectives}

\author{L. Leuzzi} 
\affiliation{Statistical Mechanics and Complexity Center (SMC) - INFM
  - CNR, Dipartimento di Fisica, Universit\`a Sapienza di Roma,
  P.le Aldo Moro 2, I-00185 Roma, Italy}

\begin{abstract}
In this paper we present a short survey on the concept of effective temperature,  on its onset as a glass former vitrifies, on the various definitions in literature and their limits of applicability.
An exactly solvable  model glass is employed to compare effective temperatures among them and to set a criterion for the occurrence of a universal extra temperature in the framework of  a "two temperature thermodynamics" for off-equilibrium aging systems. It will be shown that aging in glass formers is not a sufficient requirement. As an instance, memory effects typical of glasses  are not compatible with a unique effective temperature.
Yet, a reduced range of applicability can still be established and investigated.
\end{abstract}

\maketitle

Thermodynamics was initially devised as the theory for the behavior
of energy exchanges in steam engines \cite{Carnot1824}.  
It was born and developed in the first half of the 19th
century as a new way of looking at phenomena that, in contrast with
the Newtonian mechanics approach, were not deterministic nor
predictive. Its goal was to establish the constraints imposed by
nature in the exploitation of its forces, and to control and drive
energy transformations in order to estimate the optimum performance
of a thermal machine.  The fact that the theory was later mainly
developed at equilibrium does not mean that the equilibrium hypothesis
is the fundamental issue of thermodynamics.  The
difficulties met so far in the attempt of using thermodynamics for
glasses could be simply related to the unfounded equilibrium
hypothesis.

In this paper we will  address the issue of building a
thermodynamics working also for systems out of
equilibrium, at least in the time regime where aging
and separation of timescales occurs.  In
order to do that, we will insert the time dependence of the relaxing
observables into effective thermodynamic-like parameters, checking
whether or not it is possible to synthesize the system's features into
one unique {\em effective temperature}.

This extra variable is fundamentally a quantity keeping track not
simply of the age of the system, but of its whole history, including,
e.g., the cooling rate under which the glass
has been formed \cite{Nieuwenhuizen97,Leuzzi07}.  In some cases, making use of the effective
temperature it has been actually possible to
connect, in the space of thermodynamic parameters, the liquid and the
glass phase like in a standard thermodynamic transformation \cite{Nieuwenhuizen00}.

We shall discuss  the possibility
that, within a yet unknown class of systems, under a fixed dynamical
protocol (e.g., cooling at a fixed rate or quenching very rapidly) the glassy state is
described by this extra state variable. 
 This relies on having,
together with a set of fast processes that are in instantaneous
equilibrium, also a set of slow modes with a much larger
characteristic timescale.  This timescale can be the age of the
system (for isolated aging systems), or else, the
inverse of the cooling rate under which the glass has been
formed.  
The slow modes can be so slow that they set out a sort of
quasi-equilibrium at some effective temperature $T_e$, 
slowly depending on time. 
 In good cases, the same
effective temperature describes a variety of different physical
phenomena on a given time-scale (some decades wide), as if the slow modes carrying the structural
relaxation were at an equilibrium in a thermal bath
at that effective temperature.  
In less lucky cases, one extra parameter is not enough to implement a thermodynamic description of the glass and other parameters can be added to this aim.
If, however, at the end of the day one ends up needing
as many effective parameters as the number of principal observables of
the system, the thermodynamic description loses completely any
character of generality, being no more than a reformulation of the dynamic
behavior of each observable (as was the case for the so-called
{\em fictive temperature} \cite{Naravanaswamy71}).

After a very concise summary of the different effective temperatures introduced in literature so far,
in order to inspect the robustness of the concept of effective temperature  we shall look at an exactly solvable glass model, where the dynamics is under control at any time and all effective temperatures can be computed in terms of the model observables.

\section{Landscape and configurational entropy of a glass former}

\begin{figure}[t!]
\includegraphics[width=0.95\columnwidth]{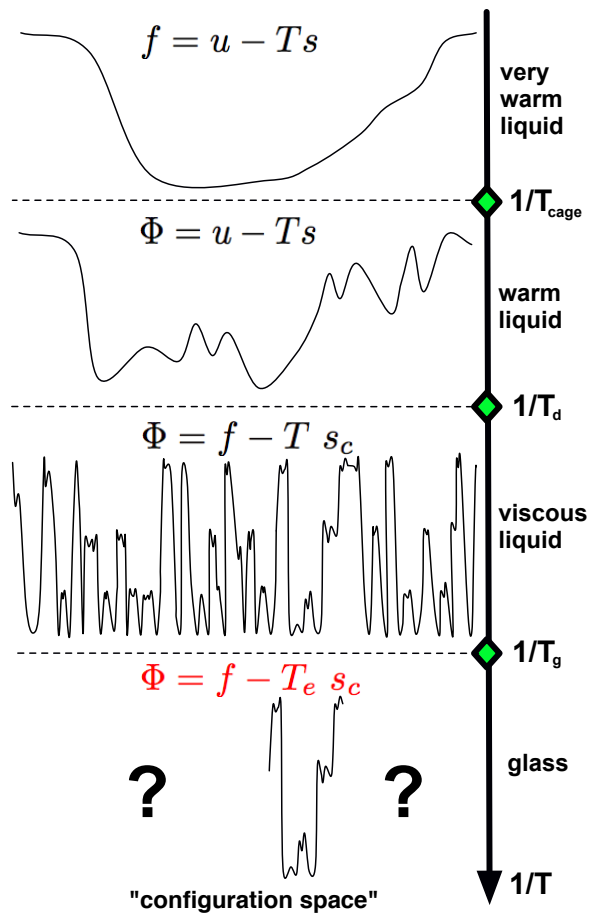}
\caption{Temperature behavior  of an observable $O$ (energy, enthalpy, volume) of a glass former through the glass transition $T_{\rm g}$ on cooling
(upper curve) and reheating (lower curve). The dashed line represents
the extrapolation to low temperature of the
relaxation values of $O$ in the liquid
phase. $T_f^{(A)}$ is the fictive temperature relative to the
relaxation of $O$ at the point $A$ in the cooling, i.e., when the system is at temperature $T_{\ell}$.
}
\label{fig:landscape}
\end{figure}

We first recall the vitrification process of a liquid glass former in a cooling procedure starting at high temperature in terms of free energy landscapes and the relative free energy vs. entropy relationships.
In Fig. \ref{fig:landscape} we show a very simplified picture of the free energy as a function of the system configurations, there drastically synthesized by a single variable on the abscissa.
Resorting to  the concept of configurational entropy, or complexity, we then put forward a first thermodynamic definition of a second   {\em effective} temperature $T_{\rm e}$, next to the heat-bath temperature $T=1/\beta$.

For a viscous liquid, the existence of processes evolving on (at least) two well separated time-scales ($\alpha$ and $\beta$ processes) can be mapped into a free energy landscape in the space of configurations with well separated basins.
Each basin yields an entropic contribution $s$  counting all configurations linked by fast processes,  and displays a local free energy: the  "single state"  free energy $f$.
The crossing of the basin barriers and the evolution
 into another basin needs a $\alpha$ process to take place. All other basins contribute to the total entropy with a {\em configurational} contribution $s_c$.
 The {\em total} free energy of the supercooled liquid is, then, written as
\begin{eqnarray}
\Phi(\beta)&=&f-T s_c(f)
\end{eqnarray}
where $\beta\Phi$ and $s_c$ are Legendre transforms of each other and
\begin{eqnarray}
\beta&=&\frac{\partial s_c}{\partial f}
\label{f:ttt1}\\
f&=&\frac{\partial \beta\Phi}{\partial \beta}\ .
\end{eqnarray}

As the temperature decreases below the glass temperature $T_{\rm g}$, the inter-basin processes are inhibited (on the
 time scales of observation, of course, operatively of the order of $\tau\approx 10^2-10^3$ s, corresponding to a viscosity larger than $10^{12}$ Pa s) and the configurational entropy contributions to the total free energy are inaccessible. From the point of view of the single vitrification experiment they disappear.  The complexity $s_c$ might then be considered as a state function and its intensive conjugated variable can be adopted as an extra thermodynamic parameter with respect to equilibrium thermodynamics:
\begin{eqnarray}
\Phi(\beta)&=&f-T_{\rm e} s_c(f)
\end{eqnarray}
with
\begin{eqnarray}
\beta_{\rm e}&=&\frac{\partial s_c}{\partial f}\\
f&=&\frac{\partial \beta_{\rm e}\Phi}{\partial \beta_{\rm e}}
\end{eqnarray}

To encode the off-equilibrium condition of the glass we, therefore, resort to a two-temperature thermodynamics (TTT), in which $T_{\rm e}$ is introduced
next to $T$.

\section{Two-temperature thermodynamics}

The introduction of an extra parameter in thermodynamics in order to describe nonequilibrium  phenomena
goes back to Reiss \cite{Reiss97} and Gutzow \cite{Gutzow95}.
An equivalent formulation was recently  put forward by Nieuwenhuizen \cite{Nieuwenhuizen97,Nieuwenhuizen98a,Nieuwenhuizen98b} in terms of the above mentioned effective temperature. One can devise a generalization of thermodynamics can be devised holding for systems 
having some components at equilibrium at a temperature $T$ and some others at $T_{\rm e}$. 
A detailed overview of the theory and its developments and applications  can be found in Ref.  \cite{Leuzzi07}. Here  we  show, anyway,  a 
summary of basic laws and relationships of TTT.

In the TTT framework,
the 
first and second law (in the Clausius inequality formulation) can be expressed as:
\begin{eqnarray}
dU&=&\delta Q+\delta W = TdS+T_{\rm e}dS_c-pdV
\\
\delta Q &\equiv& T dS + T_{\rm e }dS_c \leq TdS_{\rm tot}=T d(S+S_c)
\\
 &&\leftrightarrow T_{\rm e}-T dS_c\leq 0
\end{eqnarray}

The free energy potentials read
\begin{eqnarray}
\Phi&=&U-TS-T_{\rm e} \qquad\quad\mbox{Helmoltz}
\\\gamma&=& U -TS-T_{\rm e}S_c+pV\qquad \mbox{Gibbs}
 \end{eqnarray}
 
 The generalized Maxwell relation between entropy changes and volume changes becomes
 \begin{equation}
 \frac{\partial V}{\partial T}\Bigr|_{p}+\frac{\partial S}{\partial p}\Bigr|_{T}=
 \frac{\partial S_c}{\partial T}\Bigr|_{p}\frac{\partial T_{\rm e}}{\partial p}\Bigr|_{T}
 -\frac{\partial S_c}{\partial p}\Bigr|_{T}\frac{\partial T_{\rm e}}{\partial T}\Bigr|_{p} 
 \end{equation}
 were the r.h.s. is zero at equilibrium.
 
 Eventually, we can express in the TTT formulation the Ehrenfest-Keesom (EK) relations characterizing
 second 
 order phase transitions \footnote{The glass transition at $T_{\rm g}$
 resembles a smeared second order phase transition with  sharp changes in specific heat, compressibility and heat capacity but continuous in energy, enthalpy and volume (no latent heat).}
  and their ratio, called 
    Prigogine-Defay ratio $\Pi$.
 The mechanical, or first, EK relationship, between the jumps in thermal expansivity - $\Delta\alpha$ -  and 
in specific heat  - $\Delta C_p$ -  does not involve thermodynamics and, therefore, being out of thermodynamic equilibrium does not affect it. It is the same as at equilibrium:
 \begin{equation}
 \frac{\Delta\alpha}{\Delta\kappa}=\frac{dp_{\rm g}}{dT}
 \end{equation}
 
On the contrary, the calorimetric, or second, EK relation displays an extra term with respect to equilibrium (the second term on the r.h.s.):
  \begin{equation}
 \frac{\Delta C_p}{T_{\rm g}V\Delta\alpha}=\frac{dp_{\rm, g}}{dT}
 +\frac{1}{V\delta \alpha}\left(1-\frac{\partial T_{\rm e}}{\partial T}\Bigr|_{p}\right)\frac{\partial S_c}{\partial T}
 \end{equation}
 
 The Prigogine-Defay ratio  is often used as an order parameter to quantify the glassiness of a system.
 In  TTT reads 
   \begin{equation}
 \Pi\equiv\frac{\Delta C_p\Delta\kappa}{TV\Delta\alpha}=1+\frac{1}{V\Delta\alpha}\left(
1-\frac{\partial T_{\rm e}}{\partial T}\Bigr|_{p} \right)\frac{d S_c}{d p}\ .
\label{f:PDR}
 \end{equation}
 The common belief is that, while at equilibrium $\Pi=1$ a glass should yield $\Pi>1$. This is apparently confirmed by experiments\cite{Leuzzi07}.  $\Pi$ would be a measure of how far from equilibrium  the system is.
From Eq. (\ref{f:PDR}), however, one can see that the off-equilibrium extra term (second on the r.h.s.)  can actually take {\em any} value (not just positive) and $\Pi$ has, consequently, no special meaning as an order parameter. 
A possible explanation of the discrepancy of this results with previous experiments is that,
sometimes, compressibility is measured with techniques based on equilibrium, rather than simply using its definition  $\kappa=-\partial\log V/\partial p|_T$. This can lead to a different value of $\Delta \kappa$ at the glass transition and, consequently to a different value of $\Pi$.
For example, in the atactic polystyrene glass of Rehage and Oels \cite{Rehage77} where volume data for different, nearby, $p$ and $T$ were also recorded, Nieuwenhuizen \cite{Nieuwenhuizen97} could recalculate the ratio finding $\Pi=0.79$, rather than $\Pi=1.09$ as measured by the authors in the original paper.

\subsection{The effective temperatures}
Looking at the literature of the last 60 years, since the work of Tool \cite{Tool46} the idea of effective, or fictive, temperature has been appearing and returning both in experiments and in theories. In the following we present a hopefully comprehensive list of the most recent entries.
Apart from the definition above, Eq. (\ref{f:ttt1}) as the intensive parameter conjugated to the configurational entropy, the following effective temperatures are introduced.

{\em Gibbs-like effective temperature}.\\
We can consider slow non-equilibrium processes as if at equilibrium at a temperature different from the heat-bath temperature. Or, softening this statement, we can consider  them as if belonging to an ensemble at equilibrium at a different temperature {\em and} different values of the external fields acting on the system. In this case the effective temperature (and effective fields)
can be introduced
 through the modified Gibbs measure \cite{Nieuwenhuizen00}
\begin{equation}
\mu(C)\sim \exp\left\{-\beta_{\rm e}{\cal H}_{\rm e}[C;T, {\bm h}_{\rm e}]\right\} \ .
\label{f:ttt2}
\end{equation}

{\em Fictive temperature}.\\
 The fictive temperature is defined as the temperature at which the glass would have been if the ordering behavior on relaxation in the liquid phase would have continued below $T_{\rm g}$ \cite{Naravanaswamy71}.
\begin{figure}[b!]
\includegraphics[width=0.95\columnwidth]{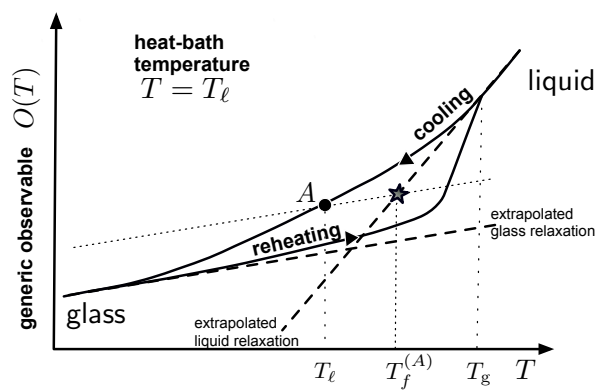}
\caption{Temperature behavior  of an observable $O$ (energy, enthalpy, volume) of a glass former through the glass transition $T_{\rm g}$ on cooling
(upper curve) and reheating (lower curve). The dashed line represents
the extrapolation to low temperature of the
relaxation values of $O$ in the liquid
phase. $T_f^{(A)}$ is the fictive temperature relative to the
relaxation of $O$ at the point $A$ in the cooling, i.e., when the system is at temperature $T_{\ell}$.
}
\label{fig:fictive}
\end{figure}
To be quantitative we can take into account a cooling heating experiment accross the glass transition, as sketched in  Fig. \ref{fig:fictive}, looking at one particular observable $O$ (energy, enthalpy, volume,...).
For high $T$ the $O(T)$ line is the same on cooling and heating. If no dynamic arrest would take place (nor crystallization) the liquid behavior would continue to low $T$, as represented by the dashed line.
For low $T$ the glass former is solid and stable and the $O(T)$ curve on cooling and heating is also
reversible. The fictive temperature $T_{\rm f}$ is related to the temperature derivatives of $O$ deep in the glass and in the liquid (equilibrium) phase as \cite{Moynihan76}:
\begin{equation}
\frac{dT_{\rm f}}{dT}\Bigr|_{T_\ell}=
\left[\frac{d O}{dT}-\left(\frac{\partial O}{\partial T}
\right)_{\rm g}
\right]\left[\left(\frac{\partial O}{\partial T}\right)_{\rm eq}
-\left(\frac{\partial O}{\partial T}\right)_{\rm g}
\right]^{-1}
\end{equation}
If we look at the value of $O$ at $T=T_\ell$ in the glassy phase (point $A$) its fictive temperature is the abscissa of the crossing point between the liquid extrapolation line and the line passing through $A$ with the slope of the extrapolated glass relaxation line. As we probe lower temperature, deep in the glass phase, the fictive temperature equals a limit value determined by the crossing point of the extrapolated liquid and glass relaxation lines (Fig. \ref{fig:fictive}).
As it has been recognized already 30 years ago, cf., e.g., Ref.  \cite{Moynihan76}, however, $T_{\rm f}$ depends on the observable considered and is not a real thermodynamic parameter.

{\em Dynamic transition rate effective temperature}.\\
In systems where the transition rate  in the dynamics between two configurations whose energy difference is $\Delta E$ is known, the effective temperature can be introduced as:
\begin{equation}
\frac{W(\Delta E)}{W(-\Delta E)}\sim e^{\beta_{\rm e}\Delta E}
\label{f:dtr}
\end{equation}
where Eq. (\ref{f:ttt1}) is used to obtain the exponent \cite{Leuzzi02b,Crisanti03}.

{\em Fluctuation-dissipation ratio (FDR).}\\
The fluctuation-dissipation theorem connects the time correlation  $C(t,t_{\rm w})$ with the response  $G(t,t_{\rm w})$  at time $t$ to a small perturbation at time $t_{\rm w}$. When a system is  out-of-equilibrium  the hypothesis of the theorem are not satisfied but we can generalize the relation and define  an effective temperature  as \cite{Cugliandolo93}
\begin{equation}
T_{\rm e}(t_{\rm w}) = \left\{\begin{array}{l} \frac{\partial_{t_{\rm w}}C(t,t_{\rm w})}{G(t,t_{\rm w})} \ ,
\\
\\
\frac{C(t_{\rm w},t_{\rm w})-C(t,t_{\rm w})}{\chi(t,t_{\rm w})}\ ,
\\
\\
\frac{\pi\omega S(\omega,t_{\rm w})}{\chi''(\omega,t_{\rm w})}\ ,
\end{array}\right.
\label{f:fdr}
\end{equation}
where $\chi$ is the integrated response, or susceptibility, $S$ is the spectral density (the Fourier transform of the correlator) and $\chi''$ the loss function, i.e., the imaginary part of the Fourier transform of the time dependent susceptibility. 
We stress that  the three different formulations, trivially equal to each other at equilibrium, might yield different results out-of-equilibrium, depending on the time regime considered.

\begin{figure}
\includegraphics[width=0.95\columnwidth]{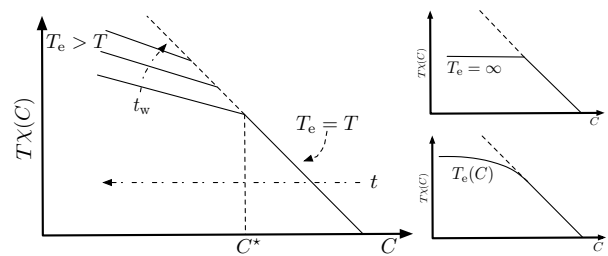}
\caption{Effective temperature as the slope of the FDR between time correlation function and 
susceptibility.
Once the measure starts (at $t=t_{\rm w}$), as time increases
the slope passes from minus one (i.e., FDR$=T$) to something
less than one (i.e., FDR$=T_{\rm e }> T$) as the correlation decreases.
 The value $C^\star$ at which the
departure from equilibrium occurs is the plateau value usually detected in slowly relaxing systems
depends on $t_{\rm w}$. In the insets  the typical 
behaviors expected for coarsening systems (top) and spin-glasses (bottom)
are shown.}
\label{fig:FDR_mf}
\end{figure}

 In Fig. \ref{fig:FDR_mf} we plot  the behavior expected  in mean-field systems for the response $\chi$ versus the correlation $C$
   for the three 
 major classes of aging systems: structural glasses, spin-glasses and coarsening systems.
 In the case of glasses it is evident that in a glass former cooled down at low temperature $T$, after a first transient in which the system relaxes as if at equilibrium, it falls out of equilibrium and the FDR is equal to a fixed quantity $T_{\rm e}$ larger than $T$. Because of aging, the point at which the glass-former 
 departs from equilibrium depends on the time waited before measurements began.

 FDR is certainly a measure of the freezing-in of the degrees of freedom due to localization, of their lack of response to external perturbations because of local constraints. 
 Whether  the FDR computed in mean-field models is a reliable measure of the fall out of equilibrium in  realistic aging systems, however, is still a contradictory subject. At least,  
for what concerns the subclass of coarsening systems \cite{Lippiello00, Godreche00, Fielding02, Nicodemi99, Crisanti00, Mayer05}.  We will not consider them here, anyway,  and we refer to Ref. \cite{Leuzzi07} and references therein for a discussion on the inadequacy of an effective temperature description of the aging dynamics in this particular case. 

Focusing on glasses there have been many numerical simulations of computer models
confirming the onset of an effective temperature as the system  vitrifies.
\cite{Barrat99,DiLeonardo00,Crisanti00,Sciortino01,Berthier02,Grigera04}.
Experiments detecting a FDR temperature behaving like the slope of the $\chi(C)$ curves of Fig. \ref{fig:FDR_mf} are, instead, very rare. To our knowledge only  a single work of the kind exist, for spin-glasses \cite{Herisson02}.

{\em Inherent structure temperature}.\\
Yet another effective temperature can be identified in the Potential Energy Landscape approach
 where the real dynamics in configurational space through the  total free energy landscape is mapped onto a symbolic dynamics through minima of the potential energy landscape, assuming a one-to-one correspondence of the minima in the two landscapes \cite{Sciortino05,Leuzzi07}.
 The inherent structure effective temperature is
\begin{equation}
T_{\rm e}=\left(1+\frac{\partial f_{\rm vib}}{\partial \phi}\right)\left(\frac{\partial s_c}{\partial \phi}\right)^{-1}
\label{f:pel}
\end{equation}
where $\phi$ is the potential energy and $f_{\rm vib}$ is the contribution to the free energy of a single basin in the potential energy.

\subsection{Direct measure}

At equilibrium the  definitions above coincide and are just equivalent expressions of the heat-bath temperature.
They represent something independently measurable with a thermometer.
It is natural to wonder, then, if in some precisely characterized subclass of 
off-equilibrium systems 
\begin{enumerate}
\item
 all above definitions coincide; 
 \item the effective temperature  can be measured by some kind of thermometer.
\end{enumerate}

If $T_{\rm e}$ is a temperature there should be a way to set up 
an {\em effective thermometer} to measure it. Besides this, there should be 
a  heat flux from modes at $T_{\rm e}>T$ (e.g., because of vitrification on cooling) to modes thermalized at $T$. Furthermore,  if more non-equilibrium modes are there, each one evolving Òas if at equilibriumÓ at a different effective temperature, there must be heat exchange among them.
Eventually, processes evolving on similar time-scales should have the same effective temperature, i.e., a zeroth law of thermodynamics should hold  on fixed time windows.

In cooling silica vitrifies at around 1800-2100 K. By definition,
 the effective temperature will encode the fall out of equilibrium occurred at those high temperatures and this memory should remain also when the glass is cooled down to room temperature ($\sim 300K$).
 There will be modes that, for time-scales comparable with our observation time, are at equilibrium among them at a $T_{\rm e}\lesssim T_{\rm g}$.

Why donÕt we burn our hands when we touch a window glass? Why the temperature measured by a thermometer is the room temperature?

A reason might be that  a thermometer has to be coupled to the slow modes, carrying the structural relaxation. That is, it should have a response time comparable with the characteristic time-scales of the structural relaxation. Another possible reason might be that the thermal conductivity of slow modes decays rapidly, hindering the heat exchange with the environment.
Experimental evidence for these conjectures are, however, lacking and the theoretical study of the measurability problem of the slow modes temperature is still at a speculative level, mainly based on the study of simple glassy models \cite{Cugliandolo97, Garriga01,Kurchan05}.

Summarizing what we recalled until now, we have seen that the idea of an unique effective temperature
and the related  two temperature thermodynamic description of off-equilibrium systems suffer of some drawbacks and many uncertainties.
There are many definitions that are {\em not} always proved equivalent. 
At least in coarsening systems it can turn out to be observable dependent or even negative, or yet higher than the transition temperature (on cooling).
On glasses there is no direct measure, nor real experiment attempted or proposed. 
How universal, or better saying, how {\em  thermodynamic} 
 is, then, the concept of effective temperature?

It might be the modern analogue of the fictive temperature, that is not a temperature but a parameter characterizing the slow relaxation of a glass-former, it might just be 
 an alternative rephrasing of aging relaxation behavior.
Or, else, it might be simply  {\em less universal} than initially expected. And yet an useful tool under specified conditions. If so, under which conditions?
In order to clarify this issue,   we adopt in the following a simplified approach, exploting the features of  a dynamically facilitated exactly solvable model for a glass \cite{Leuzzi07}.
\\
\section{Exactly solvable model approach}
 We report some results on a class of models displaying the properties both of a strong and a fragile glass \cite{Leuzzi01a,Leuzzi02a}. 
In the model we will consider
 all definitions of effective temperature given in the previous sections (apart from the fictive temperature)
 can be computed explicitely \cite{Leuzzi01a,Leuzzi01b}.
Therefore, it is possible to 
 verify, in terms of the parameters of the model, what are the conditions for having a unique effective thermodynamic parameter for all observables on a given long time-scale.
 Moreover, we are able to
 analyze what happens when those conditions are not satisfied anymore, thanks to  
the introduction of a further effective field that can help encoding the slow  aging dynamics \cite{Leuzzi01a}.
Eventually we will
 check whether a typical memory feature of glasses, the Kovacs effect \cite{Kovacs63}, can be encoded in a two-temperature thermodynamic description \cite{Aquino06a,Aquino06b}.

\subsection{Model description}
We consider a set of $N$ uncoupled harmonic oscillators $x_i$, each one locally coupled to a spherical spin $s_i$:
\begin{equation}
{\cal H}[\{x_i\},\{s_i\}]=\sum_{i=1}^N\left(\frac{K}{2}x_i^2-Hx_i-Jx_is_i-Ls_i\right)
\end{equation}
with the constraint $\sum_i s_i^2=N$.
The statics is trivial and does not yield anything glassy.
If, however, we introduce {\em ad hoc} the glass-like time-scale separation between fast variables (spins) and slow variables (harmonic oscillators) and we integrate out the fast modes we obtain the
effective potential 
\begin{eqnarray}
{\cal H}_{\rm e}[\{x_i\} &=& -\frac{1}{\beta}\log \int \prod_{i=1}^N ~e^{-\beta{\cal H}[\bm{x},\bm{s}]}
\delta\left(\sum_i s_i^2-N\right)
\nonumber
\\
&=&N\Biggl\{\frac{K}{2}m_2-Hm_1-w(m_1,m_2)
\\
&&\hspace*{2cm}+\frac{T}{2}\log\left[1+\frac{T}{2w(m_1,m_2)}\right]
\Biggr\}
\nonumber
\\
 w(m_1,m_2)&\equiv& \sqrt{J^2m_2+JLm_1+L^2+T^2/4}
\\
 m_a&\equiv&\frac{\sum_i x_i^a}{N}=\frac{M_a}{N}
\end{eqnarray}

This is the free energy of the spin coordinates given a configuration of $\{x_i\}$. From the point of view of the oscillators dynamics, though, it is still a Hamiltonian, 
with some non linear terms due to the noise induced by the presence of fast processes.

The key ingredient is, then, a {\em Parallel} Monte Carlo (PMC) dynamics that, thanks to the simplicity of the model, can be implemented analitically leading to a set of integro-differential equations
for one-time ($m_1,m_2$) and two-time (correlation and response functions) observables.
The dynamics we implement on the model has been initially introduced for the Sherrington-Kirkpatrick model in spin-glass theory \cite{Bonilla96} and for the  simpler model of uncoupled oscillators \cite{Bonilla98}.
The updates $r_i$ are randomly Gaussian distributed and small ($\sim 1/\sqrt{N}$) and the dynamic protocol is synthesized as follows:
\begin{eqnarray}
&&x_i\to x_i'=x_i+\frac{r_i}{\sqrt{N}}\qquad \forall i\\
&& P(r_i)=\frac{1}{\sqrt{2\pi\Delta^2}}\exp\left(-\frac{r_i^2}{2\Delta^2}\right)
\\
&&\Delta E={\cal H}_{\rm e}[\{x_i'\}]-{\cal H}_{\rm e}[\{x_i\}]
\\
&&W(\Delta E)=\left\{
\begin{array}{c l}e^{-\beta\Delta E} &\Delta E>0\\
1&\Delta\geq 0\end{array}\right.
\end{eqnarray}
where  $\Delta E\simeq \frac{\tilde K}{2}m_2-{\tilde H}m_1$, and
\begin{eqnarray}
\tilde K=K-J^2/(w(m_1,m_2)+T/2)
\\
 \tilde H=H+JL/(w(m_1,m_2)+T/2).
 \end{eqnarray}
The energy shift between the proposed new configuration $\{x_i'\}$ and the old configuration $\{x_i\}$ is computed after {\em all} oscillator positions have been updated. This implies a global move, even though the Hamiltonian has no global interaction (apart from the spherical constraint that is, however, irrelevant) and a consequent slow dynamics at low temperatures.

This dynamically facilitated model actually  yields  all the properties typical of glasses and
can represent both strong and fragile glasses, having the advantage of being exactly solvable. Here we only show -very briefly- the equations of motion of the one-time variables and their analytic solution for long times, while for a comprehensive didactic  presentation we refer to \cite{Leuzzi07}. The dynamic PMC equations for $m_1$ and $m_2$ are
 \begin{eqnarray}
 \dot m_a(t)&=&\int_{-\infty}^\infty \!\!\! dx~W(x)~p(x|m_1(t),m_2(t))~y_a(x);\
 \nonumber \\
 &&\qquad a=1,2
 \label{f:m_a_dyn}
 \end{eqnarray}
 where
 \begin{eqnarray}
 p(x|m_1,m_2)&=&\frac{1}{\sqrt{2\pi \Delta_x}}
 \exp
 \left[
 \frac{(x-\bar x)^2}{2\Delta_x}
 \right]
 \\
\bar x= \frac{\Delta^2\tilde K}{2}&;&\Delta_x=\Delta^2\tilde K^2(\mu_1^2+\mu_2+m_0)
 \\
 y_1(x)&=&\frac{\mu_1}{\mu_2+\mu_1^2+m_0}\frac{x-\bar x}{\tilde K}
 \\
 y_2(x)&=&\frac{2}{\tilde K}\left[x+\tilde H~y_1(x)\right]
 \\
\mu_1=\frac{\tilde H}{\tilde K}-m_1
&;& 
 \mu_2\equiv m_2-m_1^2-m_0 
 \label{f:mu1mu2}
 \end{eqnarray}
 \begin{table}[b!]
 \begin{tabular}{|l|l|}
 \hline
 $\Delta^2$&$\qquad\tau_{\rm eq}$
 \\
 \hline
 \\
 constant& $\qquad\exp\left[\frac{1}{\bar\mu_2(T)}\right]=\exp\frac{A}{T}$ 
 \\
$ \propto\mu_2(t)^{1-\gamma}$&  $\qquad\exp\left[\frac{1}{\bar\mu_2(T)^\gamma}\right]=\exp\frac{A}{(T)^\gamma} $
 \\
 $ \propto\frac{\mu_2(t)+m_0}{\mu_2(t)^\gamma}$&  $\qquad\exp\left[\frac{1}{\bar\mu_2(T)^\gamma}\right]=\exp\frac{A}{(T-T_K)^\gamma} $
 \\
 \\
 \hline
 \end{tabular}
 \caption{Correspondence between variance of the MC updates distribution and relaxation time to equilibrium. The overbar denotes the equilibrium value, expressed as a function of the temperature.}
 \label{tab:uno}
 \end{table}
 The variables $\mu_1$ and $\mu_2$ are simply recombinations of $m_1$ and $m_2$ that come into hand in manipulating the equations and in identifying dynamic regimes. They represent some sort of  distance from equilibrium. 
 The constant   $m_0$  is set equal to zero when modeling strong glasses and  to a strictly positive value value if one needs to implement a Kauzmann transition and to reproduce the properties of a  fragile glass.
Depending on the form of the variance of the PMC updates and on  the value of $m_0$, we can implement both a glass with an Arrhenius relaxation time and a Vogel-Fulcher one as is shown in Tab. \ref{tab:uno}.
In the most general case, the equilibrium is signaled by $\mu_1=0$ and $\mu_2=\mbox{max} \{T/{\tilde K}, m_0\}$. If $m_0=0$ 
equilibrium can be always reached in long enough time and in the $t\to\infty$ limit the statics is 
recovered. However, if $m_0>0$ there will be a temperature below which equilibrium
 can never be reached. This is the Kauzmann temperature.
Indeed, in the latter case shown in Tab. \ref{tab:uno} a configurational constraint is set on the oscillators such that $\mu_2(t)\geq m_0>0$  and the Kauzmann temperature $T_K$ is defined as $T_K=m_0\tilde K(\mu_1=0,\mu_2=m_0)$.  This is, actually, the temperature at which the configurational entropy
 \begin{equation}
 s_c(t;T)=\frac{1}{2}\log\left[1+\frac{\mu_2(t;T)}{m_0}
 \right]
 \label{f:s_c}
 \end{equation}
  becomes zero.

\subsection{Results}

We, now, very briefly summarize basic results of the model to show that it is, indeed, a model for glasses.

The solution to Eqs. (\ref{f:m_a_dyn})-(\ref{f:mu1mu2}) for long times is
\begin{eqnarray}
\mu_2(t)\sim \left[\log (t/t_0)+c \log(\log(t/t_0))\right]^{-1/\gamma} +\bar\mu_2(T).
\label{f:mu2t}
\\
\mu_1(t)\sim\begin{array}{ll}
\mu_2(t)^2&  \mbox{Arrhenius}
\\
\mu_2(t)^{1+\gamma}& \mbox{Vogel-Fulcher, $T>T_K$}
\end{array}
\end{eqnarray}
where we considered both the strong case and the fragile case for $T>T_K$. It can be noticed that $\mu_1$ decays always more rapidly than $\mu_2$. 
For what concerns two-time observables, as correlation function, $C(t,t_{\rm w})$, 
and response function, $G(t,t_{\rm w})$,  PMC equations of motion can be also formulated and solved for long times \cite{Leuzzi01a}.
Their solution is of the form 
\begin{equation}
C(t,t_{\rm w})=C(t_{\rm w},t_{\rm w})\frac{h(t)}{h(t_{\rm w})},
\end{equation}
where $h$ is called "time-sector function" \cite{Bouchaud98} and yields the $t$-behavior and the information relative to the aging of the system: $h(t)\sim t^\theta$, where $\theta=1/2$ for the strong glass case and
for the fragile case above $T_K$, whereas it depends on the model parameters below $T_K$.
 
 The Adam-Gibbs relation between the relaxation time to equilibrium (Tab. \ref{tab:uno}) and the configurational entropy, Eq. (\ref{f:s_c}) holds for this class of harmonic oscillator/spherical spin (HOSS) models, in a generalized form when $\gamma\neq 1$:
 \begin{equation}
 \tau_{\rm eq}\sim \exp\left[\frac{A}{Ts_c(T)}
 \right]^\gamma.
 \end{equation}
 
\begin{figure}
\includegraphics[width=0.95\columnwidth]{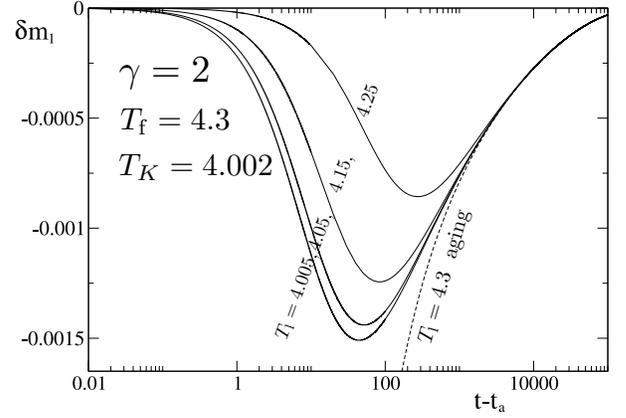}
\caption{Kovacs effect in the HOSS model with $K=J=1$, $H=L=0.1$, $m_0=5$. The dashed curve is a normal aging experiment (the final and the intermediate temperatures $T_{\rm f}$ and $T_{\rm l}$ are the same).
}
\label{fig:kovacs}
\end{figure}

 Going further, also a Kovacs protocol can be implemented on the model. 
   In place of the volume, here undetermined,
 we can take as a probe variable  the normalized distance of $m_1$ from its equilibrium value 
 \begin{equation}
 \delta m_1(t)\equiv \left[m_1(t)-\bar m\right]/\bar m_1\sim \mu_1.
 \end{equation} 
 After a quench from high temperature, we let the system evolve at a temperature $T_{\rm l}$ until  $t=t_a$ such that  $m_1(t_a;T_{\rm l})=\bar m_1(T_{\rm f})$, where $T_{\rm f}>T_{\rm l}$, i.e., we solve numerically Eq. (\ref{f:m_a_dyn}) starting from random initial conditions.
 Then we istantly heat up the system to $T=T_{\rm f}$ and we follow the PMC dynamics governed by Eqs. (\ref{f:m_a_dyn})-(\ref{f:mu1mu2}) 
with initial conditions (at $t=t_a$)  $m_1=\bar m_1(T_{\rm f})$ and $m_2=m_2(t_a;T_{\rm l})$.
The behavior of $\delta m_1(t)$ is shown in a model instance with $\gamma=2$ in Fig.  \ref{fig:kovacs}. 
The hump is clear and can be reproduced analytically, as well, combining the long time expansion leading to Eq. (\ref{f:mu2t})  and a linear, short time, expansion in $t-t_a$. For a more comprehensive treatment  the reader can refer to \cite{Aquino06a,Aquino06b}.

Even though one observable's value at $T_{\rm l}$ at a certain time  coincides with its equilibrium value 
in another heat-bath this does not imply that the system suddenly embedded in that heat-bath
 finds itself at equilibrium. Indeed, in our simple case, where the whole system evolution is represented by 
$m_1$ and $m_2$ (or $\mu_{1,2}$), the latter, not-monitored, observable is such that $m_2(t_a^+;T_{\rm f})=m_2(t_a^-;T_{\rm l})\neq
\bar m_2(T_{\rm f})$.

Now that we have looked at the glassy properties of the HOSS model we use it as a tool to 
analyze the effective temperature, after having computed explicitly the quantities presented in Eqs. (\ref{f:ttt1}), (\ref{f:ttt2}), (\ref{f:dtr})-(\ref{f:pel}).

\subsection{Effective temperatures in the HOSS model}
Defining the abbreviation 
\begin{equation}
T^\star(t)=\tilde K(\bar m_1(T),\bar m_2(T))[\mu_2(t)+m_0]
\end{equation}
 we show
the expressions of the various definitions of effective temperature in the HOSS model.
In one case we also introduce an effective field.

\begin{itemize}
\item{Conjugated to $s_c$, Eq. (\ref{f:ttt1})}
\begin{equation}
T_{\rm e}^{\rm TTT1}(t) = \left[\frac{\partial s_c}{\partial f}\Bigr|_{T}\right]^{-1} =
T^\star(t)+ O(\mu_1)
\label{f:ttt1_hoss}
\end{equation}
\item{Quasi-static (Gibbs-like), Eq. (\ref{f:ttt2})}
\begin{eqnarray}
T_{\rm e}^{\rm TTT2}(t)& =&T^\star(t) + O(\mu_1)
\label{f:Teqs}\\
H_{\rm e}(t)&=&H-\tilde K\mu_1(t)
\label{f:Heqs}
\end{eqnarray}
\item{Dynamic transition rate, Eq. (\ref{f:dtr})}
\begin{equation}
T_{\rm e}^{\rm DTR}(t) = \frac{\Delta E}{\log \left[p(\Delta E)/p(-\Delta E)\right]}
=T^\star(t) + O(\mu_1)
\label{f:dtr_hoss}
\end{equation}
\item{FDR, Eq. (\ref{f:fdr})}
\begin{eqnarray}
&&T_{{\rm e}; a,b}^{\rm FDR}(t)=\frac{\partial_{t_{\rm w}}C_{ab}(t,t_{\rm w})}{G_{ab}(t,t_{\rm w})} =T^\star(t)+
O(\mu_2^\gamma) ~\quad T>T_K
\label{f:fdr_hoss}
\nonumber \\
&&\qquad\qquad\forall ~a,b
\end{eqnarray}
\item{PEL, Eq. (\ref{f:pel})}
\begin{equation}
T_{\rm e}^{\rm PEL}(t) = \left(1+\frac{\partial f_{\rm vib}}{\partial \phi}\right)\left[\frac{\partial s_c}{\partial \phi}\right]^{-1}=  T^\star(t) + O(T\mu_2)
\label{f:pel_hoss}
\end{equation}
\end{itemize}

From the list above we can 
verify what are the 
 conditions for having a unique effective thermodynamic parameter
 for all observables on a given long time-scale. Indeed, if corrections of $O(\mu_1)$ can be neglected
 Eqs. (\ref{f:ttt1_hoss}), (\ref{f:Teqs}) and (\ref{f:dtr_hoss}) coincide.
  Eq. (\ref{f:fdr_hoss}) leads to the same result (in the leading term for long times and for $T>T_K$) only 
 if $\gamma>1$. That is, looking at Eq. (\ref{f:mu2t}), only if $\mu_2$ decays slower than $1/\log t$.
 
  We notice that Eq. (\ref{f:pel_hoss}), instead, will always give something formally different at finite $T$.  This is not surprising since the symbolic PEL dynamics is not an {\em exact} representation of the actual dynamics. The qualitative behavior is, however, the same and the numerical difference not very relevant. The corrections grow with $T$ and this implies that the inherent structure temperature is a better approximation of the effective temperature for strong glasses (where the glassy dynamics occurs for $T\approx 0$), than for fragile glasses (where we are mostly interested in the range
 $T\approx T_K>0$) \cite{Leuzzi01b}.

Exploiting Eq. (\ref{f:Heqs}) we can study what happens when the $\gamma > 1$ condition is not satisfied. 
 Admitting the introduction of a further effective field we can encode 
the slow (not slow enough for TTT) aging dynamics also when the terms of order $\mu_1$ are not negligible, and $\gamma <1$.
The effective field allows, then, for a thermodynamic description as the TTT description fails.
With two major drawbaks, though:
\\

\begin{enumerate}
\item In the quasi-static approach with a Gibbs-like
effective measure we actually introduce as many effective parameters (temperature and field) as the observables
of our system $\mu_{1,2}$, cf. Eqs. (\ref{f:Teqs}), (\ref{f:Heqs}). They come out to be a just a change of variables. Nothing is gained in this description. No universality is found.
\item the effective field has 
  no purely mechanical or thermal meaning, but is a weird mixture of the two and its physical interpretation is far from immediate.
\end{enumerate}

\subsection{Effective temperature and memory effects}
\begin{figure}[t!]
\includegraphics[width=0.95\columnwidth]{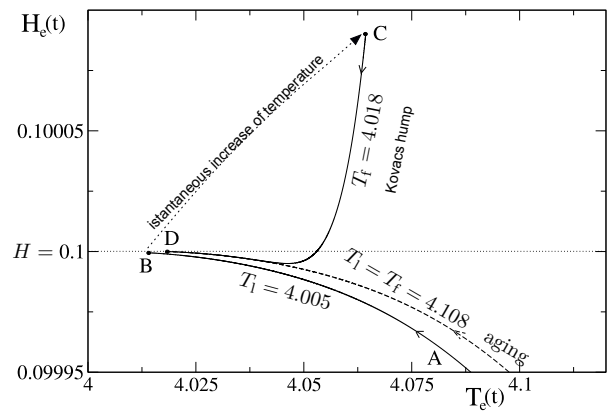}
\caption{Effective field vs. effective temperature in the Kovacs experiment on the HOSS model with parameters set as in Fig. \ref{fig:kovacs}. In particular $\lim_{t\to\infty}H_{\rm e}=H=0.1$.
The AB line is the relaxation at $T_{\rm l}=4.005$, interrupted when $\delta m_1(t=t_a)=0$.
When $T$ is increased to $T_{\rm f}=4.018$ the system ends up in the C point. The CD line represents
the  Kovacs hump and the consequent relaxation occurring at $T_{\rm f}$ in the ($T_{\rm e},H_{\rm e}$) plane. The dashed line is an aging experiment at $T=4.018$.
}
\label{fig:heff}
\end{figure}

We will now see that a unique effective temperature is not even compatible with a typical memory effect in glasses.
Indeed,
the Kovacs effect cannot be encoded in a two-temperature 
thermodynamic description.
In
Fig. \ref{fig:heff} we plot the $H_e$ vs. $T_e$ diagram of a Kovacs experiment and of an aging experiment.
One can observe that $H_{\rm e}-H$ is negligible only along the AB line, after a certain long time, that is, during a simple aging experiment (as the one plotted as a dashed line as a comparison). When we switch the temperature from $T_{\rm l}$ to $T_{\rm f}>T_{\rm l}$, however, the system's $H_{\rm e}$ jumps
to a sensitively different value than $H$ and the subsequent dynamics occurs in a regime, corresponding to the hump in Fig. \ref{fig:kovacs},
 where both $T_{\rm e}$ and $H_{\rm e}$ are necessary for a thermodynamic description.
 The very essential feature of the memory effects displayed in a Kovacs protocol is than incompatible with the TTT.

 \section{Conclusions}
 In the present work we have recalled the basic definitions of effective temperature in the literature of glassy and amorphous systems and we have  discussed the possibility of defining a self-consistent thermodynamic theory, the two-temperature thermodynamics \cite{Leuzzi07}, holding out of equilibrium on given long time-scales. The effective temperature formally plays the role of a temperature but whether this is a real one or just a parameter yielding information on the system relaxation (like its predecessor, the fictive temperature) is not known. We report on the state of the art for what concerns its measurability as a real temperature \cite{Cugliandolo97,Garriga01,Kurchan05}.
 
 In the second part of the paper we exploit the properties of a dynamically facilitated model for glassy systems, the HOSS model \cite{Leuzzi01a,Leuzzi01b,Leuzzi02a,Aquino06a,Leuzzi07}, to make an explicit  comparison between different definitions of effective temperature.
 We identify a constraint on the model parameters in order to have a working TTT. This corresponds to require not only aging and slow relaxation but a relaxation to equilibrium slower than some given function of time. Furthermore, we see that memory effects, such as the Kovacs effect, taking place in glasses in the aging regime, are not describable in the framework of a TTT.

 \acknowledgements{We thank Th.M. Nieuwenhuizen for a long-standing and fruitful collaboration on the field.}

\end{document}